# Regular solutions of Schwarzschild problem

A. LOINGER

Dipartimento di Fisica, Università di Milano

Via Celoria, 16 – 20133 Milano, Italy

**Summary.** – Through an analysis and a generalization of a profound Fock's treatment, I make conspicuous the physically essential role of those functions of the polar radial coordinate for which the expressions of the spacetime interval are everywhere regular for *all* values of the above coordinate.

**1**. – As it was emphasized by Eddington [1], an entire class **C** of expressions of the spatio-temporal interval outside a spherosymmetrical mass distribution (extended or concentrated) at rest can be obtained, *for instance*, by substituting the polar radial coordinate $r$ in the *standard* solution to Schwarzschild problem (which is due to Droste, Hilbert, and Weyl [2]) with any regular function of $r$.

Now, it is very interesting from the physical standpoint to investigate also the passage from the extended case to the limit of concentrated mass. The present writer has proved [3] that starting from the *original* Schwarzschild's solution for the field of a sphere of a homogeneous and incompressible fluid [4] the above limit gives actually the *original* Schwarzschild's solution for the field of a material point [5]. As is known, it **differs** from the standard form of solution, and can be formally obtained by substituting the $r$ in the standard formula with the expression $[r^3 + (2m)^3]^{1/3}$, where $m := GM/c^2$.

A remarkable treatment of the problem has been given by Fock [6]. He considers an extended spherosymmetrical material distribution, characterized by *any whatever* mass tensor $T_{\mu\nu}$, $(\mu,\nu = 0,1,2,3)$, and calculates the field $g_{\mu\nu}$ external to the distribution, exhibiting explicitly also the transition to the pointlike case.

I shall show that by taking suitable values for an integration constant that Fock puts equal to zero, one obtains expressions of the spatio-temporal interval for which the $g_{\mu\nu}$'s are *everywhere regular* for $r \geq 0$; accordingly, the corresponding geometrical locus of the point mass is time-like, as it is physically *necessary*. Of course, by substituting the *r* in Fock's form of solution with an arbitrary (and regular!) function $w(r)$, one has anew the mentioned class **C** of solutions to our problem. And again, the infinite instances of $w(r)$-functions for which the $g_{\mu\nu}$'s are regular everywhere for $r \geq 0$ are of paramount significance.

**2**. – Fock [6] adopts, *more solito*, spherical polar coordinates $r, \vartheta, \varphi$, and writes the $ds^2$ as follows:

(2.1) $$ds^2 = \exp[-2\Phi(r)] \ c^2 dt^2 - \exp[2\Phi(r)] \ d\sigma^2,$$

(2.2) $$d\sigma^2 := dr^2 + \rho^2(r) \ (d\vartheta^2 + \sin^2\vartheta \ d\varphi^2);$$

this $d\sigma^2$ characterizes the metric of a static three-dimensional *Bildraum* of the Einsteinian physical space, which Fock calls an "auxiliary conformal space". We shall see that outside the mass distribution this *Bildraum* is almost Euclidean (close "proximity" to Newton theory).

Then, Fock demonstrates that Einstein field equations take the following form (*where the covariance refers to the* Bildraum):

$$(2.3) \qquad \frac{1}{\rho^2}\frac{d}{dr}(\rho^2 \Phi') = -\frac{1}{2}\kappa\mu \quad ;$$

$$(2.4) \qquad \frac{1-\rho'^2}{\rho^2} = -\Phi'^2 - \kappa T_{rr} \quad ;$$

$$(2.5) \qquad -\rho\rho'' = \rho^2 \Phi'^2 - \kappa T_{\vartheta\vartheta} \quad ;$$

$$(2.6) \qquad -\rho\rho'' \sin^2\vartheta = \rho^2 \sin^2\vartheta\ \Phi'^2 - \kappa T_{\varphi\varphi} \quad ;$$

$$(2.7) \qquad T_{r\vartheta} = T_{r\varphi} = T_{\vartheta\varphi} = 0 \quad ;$$

the prime denotes differentiation with respect to $r$; $\kappa := 8\pi G/c^2$ (Fock calls the gravitational constant by the letter $\gamma$ in lieu of $G$); $\mu(r)$ represents the mass density:

$$(2.8) \qquad \mu(r) := c^2 T^{tt} + T_{rr} + \frac{1}{\rho^2}T_{\vartheta\vartheta} + \frac{1}{\rho^2 \sin^2\vartheta}T_{\varphi\varphi} \quad .$$

Obviously, outside the sphere we have $T_{rr} = T_{\vartheta\vartheta} = T_{\varphi\varphi} = 0$, and $\mu = 0$.

Integrating the equation

$$(2.3') \qquad \frac{1}{\rho^2}\frac{d}{dr}(\rho^2 \Phi') = 0$$

we obtain

$$(2.9) \qquad \rho^2 \Phi' = -m \quad ,$$

where $m$ is an integration constant (denoted with $\alpha$ by Fock). One sees easily from eq. (2.3) that $m \equiv GM/c^2$, with

$$(2.10) \qquad M := 4\pi \int_{r=0}^{r=a} \mu(r)\rho^2(r)dr \quad ,$$

if $r = a$ is the value of the radial coordinate of the points of the spherical surface.

From eqs. (2.3') and (2.4) – with $T_{rr} = 0$ –, we have

(2.11) $$\rho\rho' = (\rho^2 + m^2)^{1/2} \; ,$$

and differentiating with respect to $r$:

(2.12) $$\rho\rho'' + \rho'^2 = 1 \; ,$$

which shows that also eqs. (2.5) and (2.6) are satisfied if $T_{\vartheta\vartheta} = T_{\varphi\varphi} = 0$.

Strictly speaking, formula (2.10) gives the mass of an extended distribution, but it is clear that with suitable limits $a \to 0$, $\rho^2(r) \to 0$, $\mu(r) \to +\infty$, it gives also the mass of a material point.

The solution to eq. (2.11) is

(2.13) $$r + k = (\rho^2 + m^2)^{1/2} \; ,$$

where $k$ is an integration constant, which Fock sets equal to zero: for simplicity's sake. *Since $\rho$ must be real and different from zero*, we have the **inviolable condition**

(2.14) $$r + k > m \; ,$$

and, *outside* the mass (extended or pointlike)

(2.15) $$d\sigma^2 = dr^2 + [(r+k)^2 - m^2](d\vartheta^2 + \sin^2\vartheta \; d\varphi^2) \; .$$

Integrating eq. (2.9) with the boundary condition $\Phi(+\infty) = 0$:

(2.16) $$\Phi(r) = \frac{1}{2} \ln \frac{r+k+m}{r+k-m} \; ;$$

$c^2 \Phi(r)$ gives in the conformal *Bildraum* the (almost Newtonian) potential outside a sphere of radius $m - k$ and mass $M$. For sufficiently large values of $r$, $c^2 \Phi(r)$ gives the Newtonian potential $GM/r$.

From eq. (2.16):

(2.17) $$\exp[-2\Phi(r)] = \frac{r+k-m}{r+k+m} \; ;$$

in conclusion:

(2.18) $$ds^2 = \frac{r+k-m}{r+k+m} c^2 dt^2 - \frac{r+k+m}{r+k-m} d\sigma^2 \ .$$

For $k = 0$ we have Fock's form of solution; for $k = -m$ the standard form; for $k = m$ Brillouin's form [7]. For $k > m$ – e.g. for $k = (1+\varepsilon^2)m$ – we have *forms which hold for* $r \geq 0$: their metric tensors are regular everywhere – and consequently, the geometrical locus of the mass point is time-like, *as physics requires*. On the contrary, the analogous locus for the standard and Fock's formulae is space-like (a physical absurdity), and the locus for Brillouin's and Schwarzschild's formulae is light-like. Let us observe explicitly that eq. (2.16) for $k = -m$ gives in the *Bildraum* the gravitational potential outside a sphere of radius $2m$ and mass $M$.

**3**. – It is clear that Fock's procedure remains valid if we make the following substitution:

(3.1) $$r + k \to w(r) \ ; \ w(r) > m \ ,$$

where $w(r)$ is any regular function of $r$, which satisfies possibly the condition that its $ds^2$ is Minkowskian at the spatial infinity.

If we put in (2.18)

(3.2) $$r + k \to w(r) \equiv [r^3 + (2m)^3]^{1/3} - m \ ,$$

we obtain the original Schwarzschild's form of solution [5], which is regular for $r > 0$.

Setting

(3.3) $$r + k \to w(r) \equiv \left(1 + \frac{m}{2r}\right)^2 r - m \ ,$$

we have the solution in the isotropic system [8], here $w(r) > m$ implies $r \neq m/2$, and the locus of the mass point is time-like. However, for a point at rest we have:

$$(3.4) \qquad \frac{\mathrm{d}s}{c} = \frac{r - m/2}{r + m/2} \mathrm{d}t \quad ;$$

accordingly, we must discard the spatial region $r < m/2$ because in this region the laboratory time $t$ and the proper time $\mathrm{d}s/c$ of the above point run in *opposite* directions [9].

*Conclusion*. − The infinite forms of solution for which the spacetime interval is *regular* for $r \geq 0$ are particularly significant from the physical point of view because they satisfy *all* the criteria that are required by a sound physical interpretation. Their mere existence demonstrates the vanity of the theoretical attempts to discover new entities and new phenomena by privileging (with a misinterpretation of a well-known Birkhoff's theorem) forms of solution that are *not* regular everywhere − typically, the standard form −, and by disregarding the fact that they are *not* really valid for *all* values of the radial coordinate [10], [11].

APPENDIX

At p.140 of Weyl's fundamental memoir quoted in [9], we find a suggestive investigation concerning an **axially** symmetrical solution to the above problem of the Einstein field of a material point. For brevity, I sum up here only the essential results.

Weyl translates the isotropic expression of the $\mathrm{d}s^2$ into a "canonical" system of *cylindrical* coordinates $r_*, \vartheta_*, z_*$:

$$(A.1) \qquad \mathrm{d}s^2 = fc^2\mathrm{d}t^2 - h(\mathrm{d}z_*^2 + \mathrm{d}r_*^2) - \frac{r_*^2 \mathrm{d}\vartheta_*^2}{f} \quad ,$$

where (our *m* coincides with Weyl's *a*):

$$(A.2) \qquad f := \frac{r_1 + r_2 - 2m}{r_1 + r_2 + 2m} \quad ; \quad h := \frac{(r_1 + r_2 + 2m)^2}{4 r_1 r_2} \quad ,$$

$$(A.3) \qquad r_1 := [r_*^2 + (z_* - m)^2]^{1/2} \quad ; \quad r_2 := [r_*^2 + (z_* + m)^2]^{1/2} \quad ,$$

remark that $f \geq 0$ and $h > 0$ for *all* values of $r_1$ and $r_2$; $f = 0$ means $r_1 + r_2 = 2m$; for $r_1 \to 0$ or, alternatively, $r_2 \to 0$, we have $h \to +\infty$; the geometrical locus of *each* point of the rectilinear segment $(r_* = 0; -m \leq z_* \leq +m)$ is light-like.

It is quite evident that the simple existence of form (A.1) strengthens expressively the conclusion of sect.**3**: indeed, (A.1) does not allow the postulation of new physical objects like the BH's; the *globe* $r = 2m$ of the standard form has become a *line segment*. Moreover, Weyl demonstrates that, in the Euclidean "canonical" *Bildraum* of the Einsteinian physical space, $\psi := c^2 \ln \sqrt{f}$ represents the Newtonian potential produced by the above line segment, endowed with a uniformly distributed mass $M = c^2 m / G$. Both in Fock's and Weyl's treatments the physical role of the respective *Bildräume* is very significant [12].

[12] See further Levi-Civita's beautiful papers entitled "$ds^2$ einsteiniani in campi newtoniani": LEVI-CIVITA T., *Rend. Acc. Lincei*, vols. **XXVI**$_2$ ÷ **XXVIII**$_1$ ($1917_2$÷$1919_1$), "Nove Note"; now also in *Opere Matematiche – Memorie e Note*, vol. **IV** (Zanichelli, Bologna) 1960, p.89 *seqq*.; note VIII gives a refinement of a Weyl's argument in memoir [9].